\documentclass[fleqn,12pt,oneside]{article}
\usepackage{espcrc1}
\usepackage{amsmath}
\usepackage[dvips]{graphicx}
\input epsf
\title{Unitarity constraints and role of geometrical effects}
\author{\underline{S. M. Troshin},
 N. E. Tyurin\address{
 Institute for High Energy Physics,\\
 Protvino, Moscow Region,\\ 142280, Russia}}
\begin{document}
\maketitle
\begin{abstract}

Unitarity and geometrical effects are
discussed for
 photon--photon scattering.

\end{abstract}

\section*{Introduction}
In \cite{eur02} we have studied  limitations off--shell unitarity
provides for the $\gamma^* p$--total
 cross-sections and geometrical effects in the
  energy dependence of $\sigma^{tot}_{\gamma^* p}$.
It was shown that unitarity by itself does not lead
to the saturation at $x\to 0$, i.e.  slow down of the
power-like energy dependence of $\sigma^{tot}_{\gamma^* p}$ and transition
to the energy behavior  consistent with the Froissart--Martin bound valid
for the on--shell  scattering.
In particular,  the $Q^2$--dependence of the constituent
quark
 interaction radius  rising with the
virtuality $Q^2$  leads to an asymptotical result:
\[ \sigma^{tot}_{\gamma^* p}\sim (W^2)^{\lambda(Q^2)},
\]
 where $\lambda(Q^2)$ will not depend on virtuality  at
large values of $Q^2$.

Here we consider similar problems for
the $\gamma^*\gamma^*$--scattering.
The process of the virtual two--photon scattering is under active
study nowadays since it was expected that
the hard interaction QCD dynamics would be tested
in the most unambiguous way and
unitarity would play a minor role there \cite{brod}.
 On the other side there are
model approaches which  impose saturation in the
$\gamma^*\gamma^*$--interactions, i.e. limitations inherited from
a hadron--hadron on--shell scattering are
extended to this case (cf. \cite{moty}  and references
therein). Such approaches also consider the role of unitarity
for the wide range values of virtualities.
Available high-energy experimental data obtained at LEP
are  not  restrictive since the data are extracted
using Monte-Carlo generators \cite{lep}.

\section{Unitarity and total cross--sections of
real and virtual  $\gamma\gamma$--interactions}
Extension of the $U$--matrix unitarization for the off-shell
scattering was considered in \cite{ttpre,eur02}. To apply an extended
unitarity to DIS at small $x$ there was supposed that
the virtual  photon fluctuates into a quark--antiquark
pair $q \bar q$ and this pair  was treated as an effective
 virtual  vector meson state
in the processes with small   $x$.
This effective virtual meson  interacts then with a hadron. We
considered  a single effective vector meson field.
To treat $\gamma^*\gamma^*$--scattering we introduce the
amplitude $F^{**}_{**}(s,t,Q_1^2,Q_2^2)$
when both initial and final mesons  are off mass and $F_*^{*}(s,t,Q_1^2,Q_2^2)$
when only initial mesons are off mass shell:
\begin{equation}
V^*+V^*  \to  V^*+V^*,\quad \mbox{and}\quad
V^*+V^* \to  V+V
\end{equation}
The  amplitude $F(s,t)$ describes  the on--shell $VV$ scattering.

The unitarity  for the amplitudes $F_{**}^{**}$ and
$F_*^*$  in  impact
parameter representation at high energies   relates them
 in the following way
\begin{equation}\label{offs}
\mbox{Im} F_{**}^{**}(s,b,Q_1^2,Q_2^2) =  |F_{*}^{*}(s,b,Q_1^2,Q_2^2)|^2
+\eta_{**}^{**}(s,b,Q_1^2,Q_2^2),
\end{equation}
where $\eta_{**}^{**}(s,b,Q^2)$ is the contribution to the unitarity of
the many--particle intermediate on--shell states.
The solution of the off--shell unitarity relations
 has a simple form in the impact
parameter representation \cite{ttpre}:
\begin{eqnarray}
F_{**}^{**}(s,b,Q_1^2,Q_2^2) & = & U_{**}^{**}(s,b,Q_1^2,Q_2)
+iU_*^{*}(s,b,Q_1^2,Q^2_2)F_*^{*}(s,b,Q_1^2,Q^2_2)\nonumber\\
F_*^{*}(s,b,Q_1^2,Q_2) & = & U_*^{*}(s,b,Q_1^2,Q_2^2)
+iU_*^{*}(s,b,Q_1^2,Q_2^2)F^{}(s,b).\label{es}
\end{eqnarray}
The solution  of this system has a simple form when the following
factorization  is imposed
\begin{equation}
[U_*^{*}(s,b,Q_1^2,Q_2^2)]^2-U_{**}^{**}(s,b,Q_1^2,Q_2^2)U(s,b)=0.\label{zr}
\end{equation}
Similar factorization was implemented in \cite{eur02}. It should be noted
that these factorization formulas have been implied
at the level of  an input dynamical
quantities and they do not lead to the factorization for the
corresponding total cross--sections. Breaking of this factorization
is one of the consequences
of the unitarity.

Eq. (\ref{zr}) implies the following representation for the
functions $U_{**}^{**}(s,b,Q_1^2,Q_2^2)$ and $U_*^{*}(s,b,Q_1^2,Q_2^2)$:
\begin{eqnarray}
 U_{**}^{**}(s,b,Q_1^2,Q_2^2) & = & \omega ^2(s,b,Q_1^2,Q_2^2)U(s,b)\nonumber\\
U_*^{*}(s,b,Q_1^2,Q_2^2)& = & \omega(s,b,Q_1^2,Q_2^2)U(s,b).\label{fct}
\end{eqnarray}
 It will be evident in the following  that this factorization,
 in particular, is valid in the
off--shell extension of the chiral quark model for the $U$--matrix
which we will consider further.  The amplitudes
$F^*_*$ and $F_{**}^{**}$ then can be written in the form
\begin{eqnarray}
F_{*}^{*}(s,b,Q_1^2,Q_2^2) & = & \frac{U_{*}^{*}(s,b,Q_1^2,Q_2^2)}{1-iU(s,b)}=
\omega(s,b,Q_1^2,Q_2^2)\frac{U(s,b)}{1-iU(s,b)}
\label{vrq}\\
 F_{**}^{**}(s,b,Q_1^2,Q_2^2)& = & \frac{U_{**}^{**}(s,b,Q_1^2,Q_2^2)}{1-iU(s,b)}=
\omega^2(s,b,Q_1^2,Q_2^2)\frac{U(s,b)}{1-iU(s,b)} \label{vr}
\end{eqnarray}
and unitarity  does constraint
 the magnitudes of the above amplitudes by
unity.

The   off--shell extension of the model for
hadron scattering \cite{csn}, which uses the notions
 of chiral quark
models was developed in \cite{eur02}. The further extension for
 the case   when  both
of the colliding particles (vector mesons)  are off
mass shell  the corresponding $U$--matrix, i.e.
$U_{**}^{**}(s,b,Q_1^2,Q_2^2)$ should be represented as the
 product
\begin{equation} U_{**}^{**}(s,b,Q_1^2, Q_2^2)\,=\, \prod^{n_{V_1}}_{i=1}\,
\langle f_{Q^*_i}(s,b,Q_1^2) \rangle \prod^{n_{V_2}}_{j=1}\, \langle
f^{}_{Q^*_j}(s,b,Q_2^2) \rangle .\label{prdv}
 \end{equation}

 Factors $\langle f_{Q^*}(s,b,Q_i^2)\rangle$
 correspond to the individual quark
scattering amplitudes  smeared over the transverse position of the
constituent quark
  inside the virtual vector meson
and over the
fraction of longitudinal momentum of the initial
  parent vector meson.
Under the virtual constituent quarks $Q^*$ we mean the ones
 composing the virtual
 meson.

Further steps are completely similar to the ones described in \cite{eur02},
where the introduction of the $Q^2$ dependence into the interaction radius of constituent
quark constituent quark (which in the present approach consists of a current quark
and the  cloud of quark--antiquark pairs of the different
flavors \cite{csn}) is the main issue of the model.

Dependence on virtuality $Q^2$ comes through dependence of the
intensity of the virtual constituent
 quark interaction $g(Q^2)$
and the constituent quark interaction radius $r_{Q^*}(Q^2)=\xi(Q^2)/m_{Q^*}$
 (in the on-shell limit $g(Q^2)\to g$ and
$\xi(Q^2)\to\xi$).
 The explicit functional dpendencies
for the generalized
reaction matrices $U_*^*$ and $U_{**}^{**}$
 can be written then in the form of (\ref{fct}) with
\begin{equation}\label{omegfact}
  \omega(s,b,Q_1^2,Q_2^2)=\omega(s,b,Q_1^2)\omega(s,b,Q_2^2).
\end{equation}
and
\begin{equation}\label{omeg}
  \omega(s,b,Q^2)=\frac{\langle f^{}_{Q^*}(s,b,Q^2)\rangle}
  {\langle f^{}_{Q}(s,b)\rangle},
\end{equation}
i. e. in
the high--energy limit  (for the simplicity we assume here
that  amplitudes are imaginary and all the constituent
 quarks have equal masses and parameters
$g$ and $\xi$ as well as $g(Q^2)$ and $\xi(Q^2)$
do not depend on quark flavor).
 Then  the functions
$U$, $U_*^*$ and $U_{**}^{**}$ are the following
\begin{equation}
 U(s,b)  =  ig^N\left (\frac{s}{m^2_Q}\right )^{N/2}
\exp \left [-\frac{m_QNb}{\xi}\right ] \label{usb}
\end{equation}
\[
U_*^*(s,b,Q_1^2,Q_2^2)  =  \omega (b,Q_1^2)\omega (b,Q_2^2) U(s,b),
\]
\[
 U_{**}^{**}(s,b,Q_1^2,Q_2^2)  =
 \omega^2 (b,Q_1^2)\omega^2 (b,Q_2^2) U(s,b),\label{uvv}
\]
where the function $\omega$ is an energy-independent one
and has the following dependence on $b$ and $Q^2$
\begin{equation}\label{ome}
  \omega(b,Q^2)  =
  \frac{g(Q^2)}{g}\exp \left [-\frac{m_Qb}{\bar{\xi}(Q^2)}\right ]
\end{equation}
with
\begin{equation}\label{ksi}
  \bar{\xi}(Q^2)=\frac{\xi\xi(Q^2)}{\xi-\xi(Q^2)}.
\end{equation}

 For
 the on--shell particles $\omega \to 1$ and
we then arrive  to the result obtained in \cite{ttpre} at large $s$
\begin{equation}\label{ons}
  \sigma^{tot}_{\gamma \gamma}(s)\propto\frac{\xi^2}{m_Q^2}\ln ^2
  \frac{s}{m_Q^2}.
\end{equation}

We consider further  the off-shell scattering with
$\xi(Q^2)>\xi$ and
 at large
$s$ we have
\begin{equation}\label{totv}
\sigma^{tot}_{\gamma^* \gamma}(s,Q^2)\propto G(Q^2)\left(\frac{s}{m_Q^2}
\right)^{\lambda (Q^2)}
\ln \frac{W^2}{m_Q^2},
\end{equation}
and for the $\gamma^*\gamma^*$ total cross--section
the following behavior of the
total cross--section
 at large
$s$ will take place:
\begin{equation}\label{totgg}
\sigma^{tot}_{\gamma^*\gamma^*}(s,Q_1^2,Q_2^2)
\propto G(Q_1^2)G(Q_2^2)\left(\frac{s}{m_Q^2}
\right)^{\lambda (Q_1^2)+\lambda (Q_2^2)}
\ln \frac{s}{m_Q^2},
\end{equation}
where
\begin{equation}\label{lamb}
\lambda(Q^2)=\frac{\xi(Q^2)-\xi}{\xi(Q^2)}.
\end{equation}
 Thus,
the
steep energy increase
  of $\gamma^*\gamma^*$  total cross--section $\sigma^{tot}_{\gamma^* \gamma^*}
  \sim s^{2\lambda(Q^2)}$
has been predicted at $s\to\infty$.
The $Q^2$--dependence of the total cross--sections in the model is determined by
the function $G(Q^2)$  which should be chosen  on the experimental basis.

The obtained energy dependencies have an asymptotic nature. Existing
experimental data obtained at LEP have a significant error bars
and do not show clear tendencies as it was mentioned in the Introduction.

Interactions of real and virtual photons are predicted to have significantly
different  energy dependencies for the total cross--sections.

It is interesting to perform $\gamma\gamma$--studies at higher
energies. Tagging two--photon interactions in proton-proton
collisions at LHC was considered in \cite{piotr} and it seems to be
experimentally feasible to study $\gamma\gamma$--collisions at high
energies.
\section*{Acnowledgement}
We are grateful to  E. Martynov, V. Petrov
 and A. Prokudin for the interesting discussions. One of the authors
 (S. T.) is also grateful
 to E. De Sanctis and W.-D. Nowak for the  support of his
  participation in European Workshop on the QCD Structure of the
  Nucleon, Castello Estense, Ferrara, Italy, 3-6 April 2002.

\end{document}